# Biomaterials Science

ROYAL SOCIETY
OF CHEMISTRY

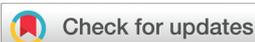
Check for updates

## Molecular origin of viscoelasticity in mineralized collagen fibrils

Mario Milazzo [ID] [a,b] Alessio David, [ID] [c] Gang Seob Jung, [a,d] Serena Danti [ID] [a,b,e] and Markus J. Buehler [ID] *[a]

Bone is mineralized tissue constituting the skeletal system, supporting and protecting the body's organs and tissues. In addition to such fundamental mechanical functions, bone also plays a remarkable role in sound conduction. From a mechanical standpoint, bone is a composite material consisting of minerals and collagen arranged in multiple hierarchical structures, with a complex anisotropic viscoelastic response, capable of transmitting and dissipating energy. At the molecular level, mineralized collagen fibrils are the basic building blocks of bone tissue, and hence, understanding bone properties down to fundamental tissue structures enables better identification of the mechanisms of structural failures and damage. While efforts have focused on the study of micro- and macro-scale viscoelasticity related to bone damage and healing based on creep, mineralized collagen has not been explored at the molecular level. We report a study that aims at systematically exploring the viscoelasticity of collagenous fibrils with different mineralization levels. We investigate the dynamic mechanical response upon cyclic and impulsive loads to observe the viscoelastic phenomena from either shear or extensional strains *via* molecular dynamics. We perform a sensitivity analysis with several key benchmarks: intrafibrillar mineralization percentage, hydration state, and external load amplitude. Our results show an increase of the dynamic moduli with an increase of the mineral percentage, pronounced at low strains. When intrafibrillar water is present, the material softens the elastic component, but considerably increases its viscosity, especially at high frequencies. This behavior is confirmed from the material response upon impulsive loads, in which water drastically reduces the relaxation times throughout the input velocity range by one order of magnitude, with respect to the dehydrated counterparts. We find that, upon transient loads, water has a major impact on the mechanics of mineralized fibrillar collagen, being able to improve the capability of the tissue to passively and effectively dissipate energy, especially after fast and high-amplitude external loads. Our study provides knowledge of bone mechanics in relation to pathologies deriving from dehydration or traumas. Moreover, these findings show the potential for being used in designing new bioinspired materials not limited to tissue engineering applications, in which passive mechanisms for dissipating energy can prevent structural failures.

## Introduction

Bone is the main constituent of the skeletal system, providing shelter to the organs and soft tissues, and allowing sound transmission and conduction for hearing.[1–3] It possesses a special hierarchical structure that has been deeply studied both with experimental and modeling tools across different length scales.[4–12] Among them, the molecular scale is the observation level that is still open to new research avenues, since the material composition and its microstructural factors determine bone properties and dynamic behavior.[13–15] Bone presents a staggered periodicity length, derived from pure collagenous tissues, called the *D*-period, which is of ≈67 nm, which is composed of two parts named overlap and gap regions.[10,16] The latter is the space where the mineral component, hydroxyapatite (HA), fills most of the interfibrillar voids.[17] As a biopolymer material working at 37 °C, collagen is subject to viscoelastic phenomena.

Viscoelastic properties of bone are crucial to understand how the tissue behaves in terms of fracture toughness,[18] failure,[18] and fatigue, among others,[19] since our daily activities

*[a]Laboratory for Atomistic and Molecular Mechanics (LAMM), USA. E-mail: mbuehler@mit.edu*
*[b]The BioRobotics Institute, Scuola Superiore Sant'Anna, Italy*
*[c]Dipartimento di Chimica, Materiali e Ingegneria Chimica "G. Natta", Politecnico di Milano, Milano, Italy*
*[d]Computational Sciences and Engineering Division, Oak Ridge National Laboratory, Oak Ridge, TN 37831, USA*
*[e]Department of Civil and Industrial Engineering, University of Pisa, Italy*

(*e.g.*, walking and running) dynamically load our bone tissues with different intensities and frequencies. Since bone has hierarchical structures, a deep knowledge of the origin of molecular level viscoelasticity may further elucidate macroscale mechanisms observed in previous studies. Specifically, bone exhibits creep and stress relaxation and is able to efficiently dissipate high-energy inputs.[20] Crack propagation and catastrophic failure of bones and tendons are the most predominant topics that have been widely investigated, but the origins of micro-damage are still under debate.[21,22] The most accredited hypotheses were related to the heterogeneity of the microstructural composition that is affected by several factors (*e.g.*, gender),[23] and atomic-scale structural changes due to shearing by accumulating the damage from cyclic loadings on tendon.[24]

A number of studies have provided a description of the mechanics of dry mineralized fibrils, highlighting the important role of HA in strengthening and toughening the collagen macromolecular chains.[10,25,26] At the same time, a previous study[26] found that collagen is the component primarily responsible for dissipating energy upon transient loads, with relaxation times that tend to decrease as fast as the increase of the input load velocity, independent of the mineralization percentage and load directionality. This feature is in good agreement with the mechanics of the tissue on the macroscale, thus giving new insights into how the physiology of the macro tissue works. Another component that influences the mechanics of bone is water, which represents about 10% of the total weight of bone, in addition to 30% of collagen and 60% of HA (Fig. 1A).[25,27,28] Researchers have demonstrated the key role of water in providing bone quality in terms of toughness[29] and in mitigating the deterioration of aging bone.[30–32]

Water is present in different forms: (i) bulk water fills the spaces of the vascular network to deliver nutrients and evacu-ate waste. It plays a negligible role in stiffening the material,[33–35] but has effects on the viscoelasticity at the macroscale;[36] (ii) structural water is present between the staggered mineralized fibrils to ensure the stabilization and assembly of bone ultrastructure;[37,38] and (iii) bound water fills the voids of the mineralized fibrils,[39,40] and is observable with nuclear magnetic resonance procedures.[34] Recently, a 2D finite element model has been proposed to investigate the effect of bound and structural water on bone mechanics, highlighting the ability of the material to tolerate external loads and microcracks, but without any specific speculation on intrinsic viscoelastic phenomena.[41] Moreover, the stiffness of the mineralized fibrils, which is monotonic with the %HA, has a linearly decreasing trend with the increase of the intrafibrillar water content, independent of the presence of minerals.[25] This interesting behavior gives the opportunity for a research question related to the effect of bound water on the viscoelastic properties of bone that, currently, have been studied mainly at the macroscale.[20,36,42–44]

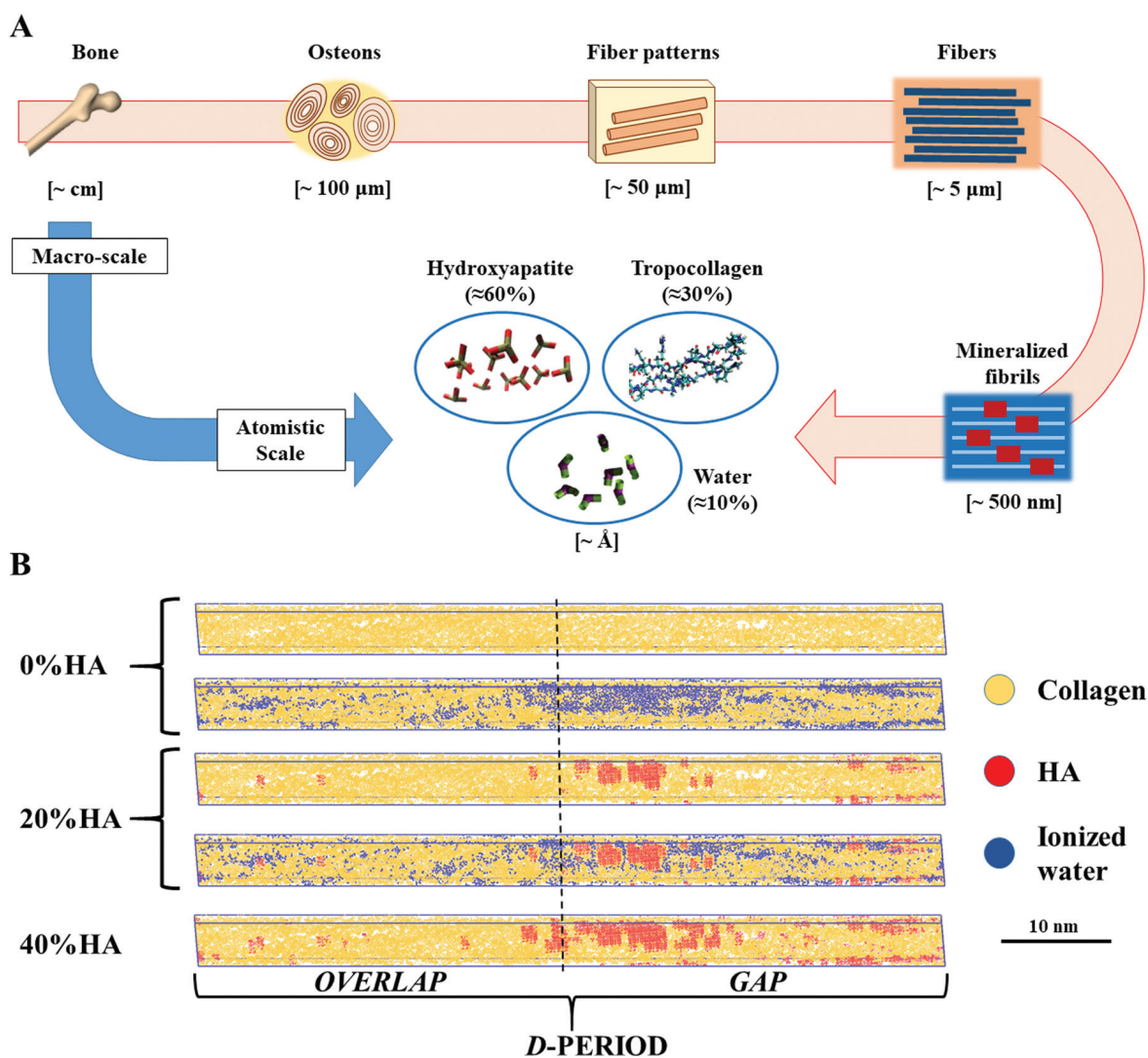

**Fig. 1** (A) Hierarchical structure of bone: from the macro tissue to the atomistic scale at which level we find tropocollagen (≈30% w/w), hydroxyapatite (≈60% w/w), and water (≈10% w/w).[26] (B) Dry and hydrated topologies employed in our studies.

In this paper, we explore the viscoelastic properties of mineralized collagen fibrils based on the different mineral and water contents at the atomic scale. Full atomistic models are utilized to understand the atomic scale behavior of dry and hydrated (also later referred to as 'wet') mineralized fibrils under cyclic shearing and impulsive loads by using molecular dynamics simulations. Understanding the mechanics of the materials upon cyclic angular loads will complement the studies on bone failure and damage in which shear deformation is more prominent than homogeneous uniaxial loads, especially at the molecular scale.

Previous works have speculated on the viscoelasticity of bone at the macroscale with experiments,[20,42,43] or with simulation tools focusing on a single collagen peptide[45,46] or the mechanics of dry mineralized fibrils upon quasi-static and impulsive loads.[26] Recently, Fielder et al. have presented a first study investigating the effect of water on the mechanical behavior of the D-period varying the %HA. They observed a decrease of Young's moduli with respect to similar dry topologies.[25] Previous studies have highlighted the role of collagen as the main responsible component for the creep phenomenon in the bone matrix at the macroscale, due to its polymeric structure that is loading-rate dependent; however, its description does not take into account the synergies with the mineral component and water.[47] Also, other studies have speculated on a possible role of water in improving the viscosity of bone, comparing the viscoelastic response of dry/wet bone upon experimental extensional cyclic loads,[43] and confirming an increase of tan δ.[20,42] To the best of our knowledge, a systematic investigation on the viscoelastic properties of mineralized fibrils, considering the role of water and %HA, has been firstly explored in this study.

Our work can also be exploited to create efficient bioinspired synthetic materials that may employ energy transmission and dissipation mechanisms, overcoming the limitations of the biological material. Moreover, from the clinical standpoint, an improved knowledge on the capability of collagen-based tissues to dissipate energy will pave the way to new treatments of pathologies or traumas, which mostly affect the elderly population,[48–50] by actively exploiting the intrinsic features of bone damage.[51]

## Results and discussion

We first conduct a study considering the D-period of the mineralized collagenous fibrils (Fig. 1B), and perform a sensitivity analysis on the elastic and viscous components of all structures under axial and tangential transient loads. Based on earlier studies on bone-like materials at the macroscale, we take advantage of the dynamic mechanical analysis (DMA, see Materials and methods section) to observe the shear viscoelastic mechanisms[42] by performing a sensitivity analysis on three different parameters: %HA, angular strain amplitude, and hydration state.

First, we investigate the properties of the dried mineralized collagen fibrils by varying the %HA. Fig. 2A–C present the viscoelastic behavior of the dry topologies with a mineral percentage ranging from 0% to 40% upon angular strain amplitudes ($\gamma_0$) of 0.017 ($\gamma_{0\,min}$) and 0.17 ($\gamma_{0\,max}$). We obtain $G'$ (storage modulus) and $G''$ as a function of $\gamma_0$ (see details in the Materials and Methods section) and describe through mathematical interpolation of the dependency of the moduli on the %HA. $G'$ appears almost constant over the frequency, independent of the mineral percentage. The frequency averaged value of $G'$ shows a clear linear trend with %HA for both $\gamma_0 s$, with a higher angular coefficient for $\gamma_{0\,min}$ (Fig. 2D):

$$
\begin{aligned}
\gamma_{0\,max}: & \quad G'(\%HA) = 577.83 + 15.02 \times \%HA, & R^2 = 0.98440 \\
\gamma_{0\,min}: & \quad G'(\%HA) = 1067.17 + 45.03 \times \%HA, & R^2 = 0.99497
\end{aligned}
\tag{1}
$$

At $\gamma_{0\,max}$, we estimate an increase of +66% from 0%HA to 20%HA, and +25% from 20%HA to 40%HA. In contrast, these increments are more pronounced at $\gamma_{0\,min}$ with +72% from 0%HA to 20%HA, and +53% from 20%HA to 40%HA (Fig. 2A and D).

The values of $G''$ show a constant behavior over the frequency range for $\gamma_{0\,min}$ (between 100 MPa and 200 MPa), but show a linear increase for $\gamma_{0\,max}$ especially for 40%HA (Fig. 2B). Therefore, we linearly fit every data series and, subsequently, linearly fit the extracted angular coefficients ($G''_m$) and intercepts ($G''_0$) of the loss modulus over the %HA (Fig. 2E and F) to evaluate the dependency of such modulus against the mineral content in the topology:

$$
\begin{aligned}
\gamma_{0\,max}: & \begin{cases} G''_m(\%HA) = 10.98 + 0.67 \times \%HA & R^2 = 0.99987 \\ G''_0(\%HA) = 212.50 + 7.07 \times \%HA \end{cases}, & R^2 = 0.99629 \\
\gamma_{0\,min}: & \begin{cases} G''_m(\%HA) = 6.30 - 0.06 \times \%HA & R^2 = 0.03894 \\ G''_0(\%HA) = 65.99 + 3.38 \times \%HA \end{cases}, & R^2 = 0.83710
\end{aligned}
\tag{2}
$$

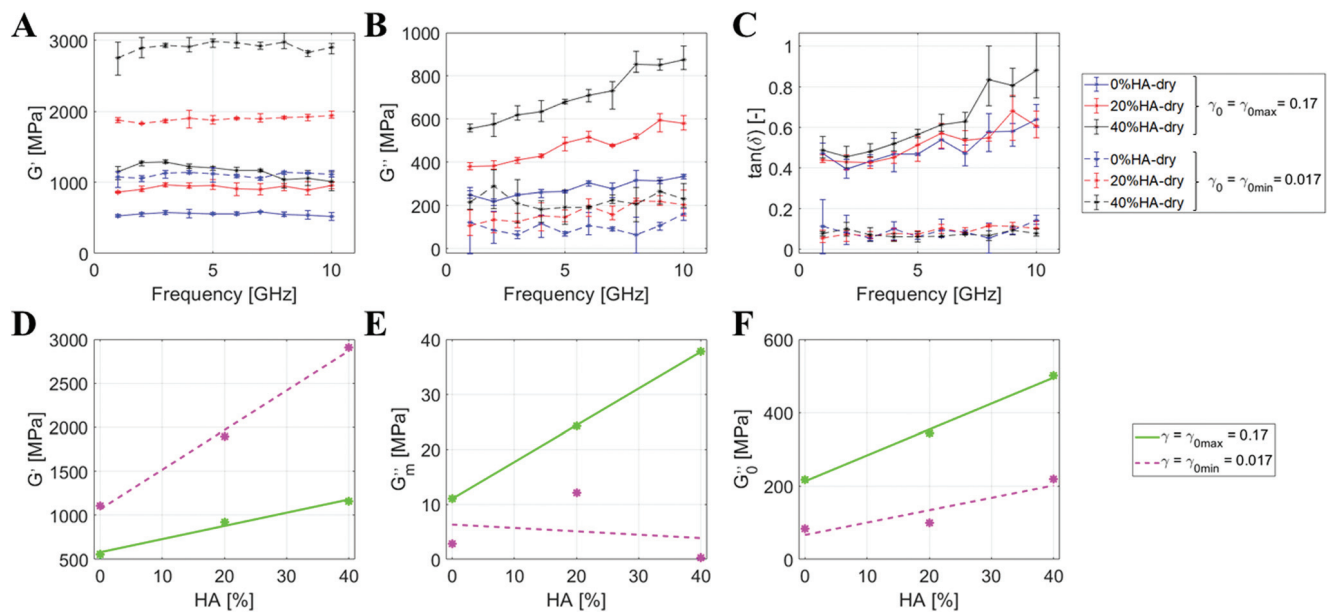

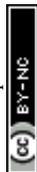

**Fig. 2** Results from the shear DMA varying the %HA and shear strain amplitude $\gamma_0$ on dry topologies. (A–C) Elastic component ($G'$), loss modulus, or viscous component ($G''$), and the lag between shear and strain (tan $\delta$) over frequency. We observe a $G'$ that is almost constant independent of the frequency for each strain amplitude. The increase of %HA leads to a stiffening of the compound that is quantitatively higher at $\gamma_0$ min. In contrast, the increase of $\gamma_0$ implies a reduction of the elastic component and a simultaneous increase of the viscosity that is more pronounced above 5 GHz. (D–F) $G'$ and $G''$ over %HA. (D) $G'$ presents linear trends over %HA with the angular coefficient that is higher at $\gamma_0$ min. (E) and (F) show the angular coefficient ($G''_m$) and the intercept ($G''_0$) for the viscous component. Also in this case, we denote linear trends in both cases. The angular coefficient at $\gamma_0$ min is very low, almost constant, while the trend for $\gamma_0$ max is markedly increasing, confirming the triggering of viscosity at high strains and frequencies. The intercepts have monotonic increasing values due to the increase of %HA.

From the intercepts, we deduce that the frequency-independent component of $G''$ is linearly dependent on the %HA, but the increase is stronger for $\gamma_0$ max, which also registers higher $G''$ values throughout the whole frequency range. The frequency dependence of $G''$, deduced from the analysis of the angular coefficients, shows a clear linear trend for $\gamma_0$ max, which is almost absent for $\gamma_0$ min, as formalized by the low $R^2$ value. The above results are consolidated through the analysis of the loss factor tan $\delta$ (Fig. 2C), which is lower and frequency-independent for $\gamma_0$ min ($\sim$0.1), while higher and frequency-dependent for $\gamma_0$ max. This shows that the tissue hardens when increasing the %HA but, at the same time, it does not show significant variations across input frequencies. In contrast, the viscous behavior is activated at high strains and is more pronounced at high frequencies, confirming the natural propensity of the tissue to efficiently dissipate high-energy inputs.

Then, we explore the difference between wet and dried systems through the same methodology. Fig. 3 shows a synoptic table comparing the DMA results for the dry/wet systems at 0%HA and 20%HA. By looking at the first column, we observe the growth for $G'$ over the %HA that is enhanced in the dry state rather than in the hydrated one ($\approx$+65% vs. 48%). $G'$ shows a frequency-independent behavior in the dry state, while we notice a marked frequency-dependent behavior at low frequency in the wet state. For both dry and wet systems, the numerical values for $G'$ at $\gamma_0$ min are about double the ones achieved at $\gamma_0$ max. In contrast, the viscous component of the material is more pronounced for the hydrated states, especially for 0%HA at high frequencies above 5 GHz with $\gamma_0$ max (Fig. 3C). $G''$ shows a frequency-dependent behavior at $\gamma_0$ max with a numerical value for tan $\delta$ up to $\approx$1.4 (Fig. 3B and C). These features are almost absent for $\gamma_0$ min, in which $G''$ and tan $\delta$ are equal to 600 and 0.5, respectively (Fig. 3E and F). Viscous behaviors are, thus, triggered by a stronger shear deformation and hydration, a trend which is also observable in Fig. 2B.

To understand the behavior under the impact of creep loading, we utilize the characterization based on the response upon axial impulses (later referred to as wave propagation – "WP"). This approach is a validated technique, exploited across different scales in earlier works, able to deliver a complete description of a material or structure with a simple broadband impulsive load.[26,45,52]

Next, we analyse the propagation and dissipation of energy along the hydrated 0%HA and 20%HA topologies from WP, and later compare the results with the outcomes from the dry materials studied in earlier works.[26] We use impulsive loads delivered with compressive displacements, investigating the material behavior across input velocities from 100 m s$^{-1}$ to 1000 m s$^{-1}$. We use wave speed and relaxation time as benchmarks to estimate the energy propagation and dissipation. Fig. 4 shows the normalized and average displacements of the $C_\alpha$ atoms over time and position from the loading interface to the fixed constraint, while Table 1 summarizes the main results after post-processing the data, comparing the outcomes with the dry topologies from earlier works.[26]

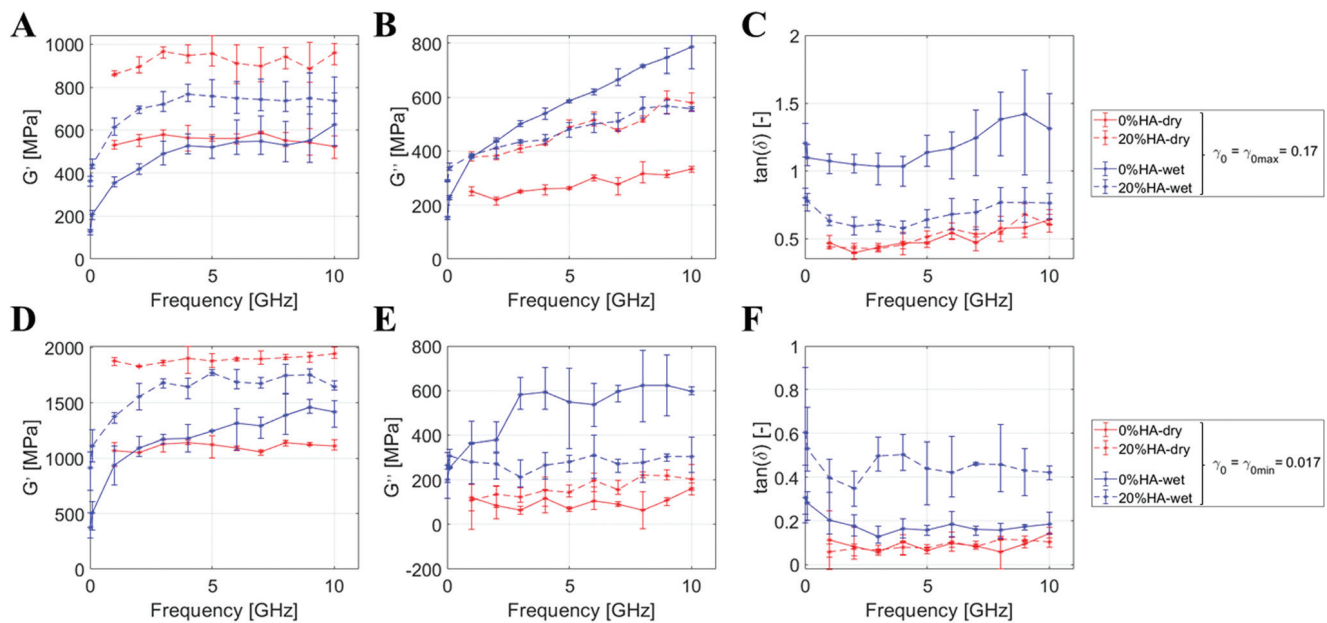

**Fig. 3** Results from the shear DMA varying the %HA and shear strain amplitude $\gamma_0$ on dry/wet topologies. Panels A–C show the elastic component ($G'$), viscous component ($G''$), and the lag between shear and strain (tan $\delta$) over frequency, at $\gamma_0 = 0.17$, while panels D–F present the results at $\gamma_0 = 0.017$. We notice a different ratio of the $G'$ between 0%HA and 20%HA for the dry and wet topologies. Specifically, hydration softens the materials and induces a monotonic increase of $G'$ up to a saturation above 3 GHz. The role of the hydration is more pronounced at 0%HA since the intrafibrillar voids are completely filled with water. It enhances the viscosity of the material, especially at high frequencies, with a growing trend above 5 GHz at high angular strains. In contrast, at lower $\gamma_0$, we observe a high viscosity that tends to a saturation over frequency.

We observe a decreasing trend of the wave speed numerical values with the increase of the input velocity (Table 1A) in the wet fibril. Specifically, 0%HA at 100 m s$^{-1}$ presents a wave speed of 3640 m s$^{-1}$ that decreases to 3039 m s$^{-1}$ at 1000 m s$^{-1}$. In contrast, 20%HA shows, on average, higher velocities from 3781 at 100 m s$^{-1}$ to 2974 m s$^{-1}$ at 1000 m s$^{-1}$. The same topologies without water molecules in the intrafibrillar voids show a general increase of the wave speeds that is more pronounced for 0%HA ($\approx$+83%) than 20%HA ($\approx$+20%).

We find the relaxation time decreases over the input velocity, although more slowly than in the dry material results, as shown in Table 1B. We observed the same trend in the hydrated one. Quantitatively, comparing the dry/hydrated states, we observe a noticeable reduction of the numerical values of about one order of magnitude, from 3.78 ps at 100 m s$^{-1}$ to 2.55 ps at 1000 m s$^{-1}$ (between 11% and 20% of the results estimated in ref. 26). These results highlight the prominent role of water in dissipating energy with relevant implications on the toughness of the tissue in the case of specific pathologies that are characterized by a reduced presence of water in bones. Our results on the relaxation times from the WP are comparable with those of the DMA upon impulsive loads. It is important to note that the definition used for this parameter is expressed by eqn (6), also used in ref. 26 and 45, and is different from the time required for the convergence of the strain upon external loads to investigate the creep in the material.[46,53] A previous study has observed a decreasing trend for the relaxation time that is stabilized at high-velocity loads in the order of 10 ps.[26] The addition of water drastically flattens the curve of one order of magnitude. This also appears in the outcomes from the DMA, in which the addition of water marks a high $G''$, especially at a frequency above 5 GHz. 20% HA is the topology that is less affected than 0%HA, since part of the intrafibrillar voids are filled by HA.[10] This outcome is mainly due to the molecular movement within the material that is enhanced by water that acts as a lubricant (*i.e.*, plasticizer) between the collagen and HA.[54] Looking at the bonds (data not shown), we observe an increment of the H-bonds for the wet topologies ($\sim$+456% at 0%HA and $\sim$+290% at 20%HA) and their cyclic breaking and recovery induces a stick-slip motion that significantly affects energy dissipation.[46,55–57]

Lastly, we estimate the mechanical properties along the fibrils by using the front-wave speed from the WP. In this case, compared to ref. 26, water fills the intrafibrillar voids (bound water), ensuring an almost full continuity to the material. From the WP, we also observe an increase of the wave speeds $\approx$83% and 20% for 0%HA and 20%HA, respectively. A direct measurement of Young's modulus is not possible using the equation:

$$E = v^2 \rho, \tag{3}$$

which relates it with the wave speed ($v$) and density ($\rho$) since the material is not homogeneous and presents different

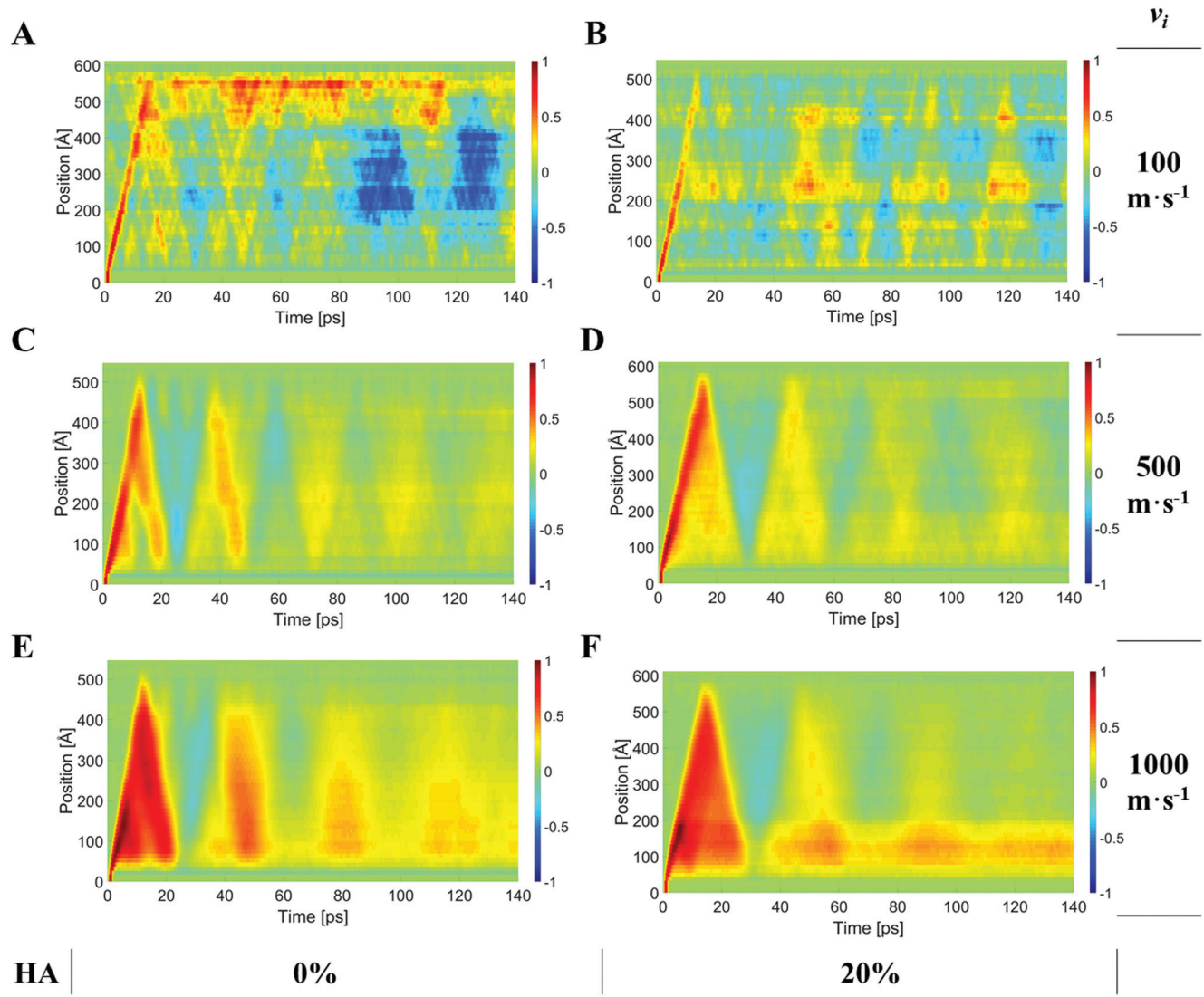

**HA** | **0%** | **20%**

**Fig. 4** Wave propagation in hydrated mineralized fibrils. Each plot shows the average displacement of the $C_\alpha$ atoms along the topologies from the loading interface to the fixed constraint over time. Panels A and B, panels C and D, and panels E and F relate to 100 m s⁻¹, 500 m s⁻¹ or 1000 m s⁻¹ input velocities ($v_i$), respectively, for the 0%HA and 20%HA wet topologies. We observe that the slope of displacement over time gets steeper (*viz.*, wave speed increases) with the input velocity and %HA. Moreover, we observe a cleaner energy transmission and faster energy at a high $v_i$, which are less disturbed by the water molecules.

**Table 1** Benchmarks for energy propagation and dissipation. A. Wave speeds estimated for the 0%HA and 20%HA dry and hydrated topologies at different input velocities, along with the average increase introduced by water in the compound. B. Average relaxation time estimated for the 0%HA and 20%HA dry and hydrated topologies at different input velocities, along with the average increase introduced by water in the compound

| A | Wave speed [m s⁻¹] | | Input velocity [m s⁻¹] | | | | | | Average increase [%] |
|---|---|---|---|---|---|---|---|---|---|
| | | | 100 | | 500 | | 1000 | | |
| | | | Dry[26] | Wet | Dry[26] | Wet | Dry[26] | Wet | |
| | % HA | 0 | 2255 | 3640 | 1817 | 3244 | 1462 | 3039 | ≈+83% |
| | | 20 | 3224 | 3781 | 2654 | 3301 | 2528 | 2974 | ≈+20% |

| B | Average relaxation time [ps] | Input velocity [m s⁻¹] | Dry[26] | Wet | Ratio [%] |
|---|---|---|---|---|---|
| | | 100 | 34.30 | 3.78 | 11 |
| | | 500 | 14.04 | 2.60 | 19 |
| | | 1000 | 12.91 | 2.55 | 20 |

phases.[58] However, the effect of water on the elasticity behavior is manifested through the DMA, in which we observe the numerical values of $G'$ that are always lower when the topologies are hydrated. This is in good agreement with the outcomes presented in ref. 25, which showed a softening due to the water component upon quasi-static loads. Moreover, the effect of water is dependent on the load frequency since it gives a monotonic growth to $G'$ up to a substantial saturation above 3 GHz. In contrast, the elasticity of the dry topologies presents a constant behavior over frequencies.

A major impact is the applicability of our study to auditory apparatus, whose components are collagen-based structures in which the frequency-dependent mechanical response is critical for the physiological functions. The tympanic membrane is a thin hydrated membrane mainly composed of collagen with fiber bundles that have different arrangements across the thickness.[59] Due to its conformation and structural organization, it filters the acoustic energy collected by the pinna and delivers a selection of the energy content to the three ossicles of the middle ear. Although the macro functions of the eardrum have been deeply studied (see for instance: ref. 52 and 60), its nanostructural behavior is still being investigated.[61] A preliminary study explained the role of water and load directionality in energy transmission and dissipation on a single collagen peptide,[62] but our outcomes on the dry/wet 0% HA topologies have the chance to elucidate the viscoelastic mechanisms at a higher hierarchy. We observe a reduced elasticity and viscosity in pure dry collagenous structures that are independent of the amplitude of the angular displacement, with numerical values almost constant across frequencies. In contrast, the addition of water promotes the viscous component of the material with a monotonic increasing trend across frequencies with numerical values that are much higher than those of the mineralized topologies. Water, indeed, enables the sliding mechanisms between collagenous fibers that are not constrained by the presence of rigid HA molecules. A similar scenario is also highlighted by the wave transmission study, in which we observe a significant reduction of the relaxation time. Hydrated collagen is the material that mostly constitutes tissues like skin, muscles and tendons that primarily receive external loads and are responsible for the first reaction or response upon external energetic inputs that may damage living bodies. Understanding the capabilities and limits of such structures may really help designing efficient tissue replacements to treat pathologies or traumas.

Furthermore, the results on mineralized fibrils help in understanding the viscoelastic mechanisms of the ossicular chain and the temporal bone, which with different mechanisms provide conductive hearing.[2,3] A deeper knowledge of healthy collagen-based structures can also pave the way to new bioinspired materials to replace and treat native tissues after traumas or pathologies.[51,63–65] With the advent of additive manufacturing and machine learning, a new generation of optimized prostheses and structures can take advantage of this study to mimic, and even improve, the behavior of native tissues for the eventual benefit of patients suffering different bone-related diseases (*e.g.*, osteogenesis imperfecta and osteoporosis) or injuries.[66–69] In view of this, new research avenues could be pursued to develop novel inks for replacing load-bearing bone parts. Hydrated constructs with peculiar viscoelastic properties could be fabricated by tuning the amount of water present in the ink.

The use of biocompatible materials has some major advantages compared to that of synthetic materials, as it promotes the inclusion of the implant in the host site by lowering the probability of rejection. Although this approach is promising, several challenges have to be considered. Firstly, the rheology of the ink highly influences the fabrication process. In fact, high viscosity promoted by water facilitates extrusion through the nozzle, but can also jeopardize the stability of the final structure. On the opposite site, the dehydration of the constructs that occurs over time may change the mechanical properties of the implant and should be prevented by efficiently incorporating water into the constructs.

## Conclusions

Bone is indisputably one of the most interesting and peculiar tissues of the human body due to its excellent mechanical properties originating from its hierarchical structures. It accomplishes different mechanical functions related to the body's support and movement, containment, and protection of vital organs, as well as sound conduction and transmission.[70] Despite the different tasks, all these applications take advantage of bone's capability in managing energy from external sources and conveying/dissipating it to prevent damage to the living tissues or to accomplish specific functions. Based on this premise, we design this work to understand bone's viscoelastic behavior from the molecular perspective, delivering a description of this specific aspect of mechanics of the material, starting from its building blocks: collagen, HA, and water.

In this study, we investigated the viscoelastic properties of dry *versus* hydrated mineralized fibrils under shear and longitudinal transient loads. From our approach, we achieve two complementary sets of results that provide the understanding of the origin of viscoelasticity in bones, and broadly, of collagen-based materials at the microscale. Even a simple axial loading is transferred as a combination of extensional and shear strain.[71–74] We unveil the origin of the tissue's capability for efficiently transferring energy with its elastic features and, simultaneously, dissipating high energy quickly from loading at its fibrillar levels.

Our study provides new knowledge on bone mechanics, highlighting new aspects at the molecular scale. The viscoelastic behaviors of mineralized fibrils help in understanding the responses of the nanostructures upon transient loads and give a new perspective on how microcracks are tolerated without causing a macroscopic failure of the tissue.[75,76] Moreover, hydration is confirmed as a key player in tissue biomechanics since it severely affects the mechanical properties, creating a synergistic effect with the solid phases of the materials and enhancing the passive mechanisms to prevent structural failure.[25,43,45,46]

# Materials and methods

## Topologies

We use the dry topologies employed in previous works to investigate the effect of mineralization on collagen fibrils with three different percentages of %HA (w%/w%): COL/HA 100/0–0%HA, representing pure collagenous structures (*e.g.*, tendons and the tympanic membrane), COL/HA 80/20–20%HA and COL/HA 60/40–40%HA. Our model is able to properly describe the intrafibrillar mineralization with the mineral component that fills the voids between collagen fibrils, mostly in the gap region.[10,62] We hydrate and randomly ionize 0%HA and 20% HA using VMD software, filling the vacuum in the simulation boxes. We exclude the 40%HA topology since the intrafibrillar voids are almost fully filled with HA.

## Force field for molecular dynamics

We perform molecular dynamics simulation using LAMMPS software (https://lammps.sandia.gov/index.html).[77] We employ an improved version of the CHARMM force field in all MD simulations, consistent with previous works.[78–80] Our parameters include hydroxyproline (HYP), which is one of the building blocks of the collagen protein structure. We use previous studies on atomistic simulations matching quantum mechanical models to provide a suitable parameter set for HYP.[81]

## Equilibration

**Full periodic topologies.** The following procedure is applied to the topologies with full periodic boundary conditions. The equilibration of the full periodic topologies includes in first instance 11 cycles, 0.5 ns each, of NVT integration at 310 K with a 1 fs time step, fixing the net momentum of all the atoms as zero. Thus, achieving the energy convergence after 5.5 ns, we use an anisotropic NPT ensemble for 6 cycles, 0.5 ns each, at 310 K and 1 atm. The final energy convergence was reached at 8.5 ns.

**Chopped topologies.** In order to study the wave propagation and energy dissipation along the structures, we chopped the bonds at the edges of the simulation box from the reference geometry, leaving the periodic conditions only along the *y*- and *z*-directions.

We preliminarily define two groups of atoms. The first group is named "bound" – BD, including the $C_\alpha$ atoms at the end of the triple helices, and the second group is labelled "mobile" – MB, including all the remaining atoms. We relax the structures with several steps to prevent abrupt structural changes from the reference. The duration of the first step is 0.1 ns with an NVE ensemble fixing the momentum of the atoms belonging to the MB group. Then, we perform two cycles of 50 ps NVT ensemble each with a Langevin thermo-stat: in the first one, we change the temperature from 10 K to 310 K, and in the second one, the temperature is kept fixed at 310 K. Afterwards, we fix the temperature of the MB group at 310 K with a Berendsen thermostat for 0.2 ns. During the equilibration, we fix the displacements and velocities of the MB group along the *y*-/*z*-axis, without any constraint along the *x*-axis.

We stabilize the system without this last constraint for 0.2 ns and, then, we use an NVT ensemble for all the atoms for 0.2 ns, fixing the temperature at 310 K. Finally, we employ an NPT ensemble for 2 ns at 310 K with a pressure constraint at 1 atm only along the *y*-/*z*-direction. A final energy convergence is reached at 2.8 ns based on RMSD, energy and pressure.

## Dynamic mechanical analysis

Dynamic mechanical analysis (DMA) is an experimental technique that aims at characterizing the viscoelastic behavior of a material upon cyclic strains.[82] In this work, we implement a computational version of the DMA by imposing a periodic angular deformation ($\gamma$) to the simulation box:

$$\gamma = \gamma_0 \sin(2\pi f t), \tag{4}$$

in which $\gamma_0$ is the deformation amplitude and $f$ is the frequency of oscillation (Fig. 5A and B).

We study the behaviour of the full periodic dry/hydrated topologies as a function of different features: the deformation frequency (*viz.*, 1 GHz–10 GHz), amplitude (0.017 rad and 0.17 rad, corresponding to the $\varepsilon_{xy}$ strain of the top face of the box of 1% and 10% of the total length of the box), mineralization percentage (0%HA–40%HA) and hydration. Following earlier studies, we also investigate the behavior of wet structures at lower frequencies (0.01 GHz and 0.1 GHz). We do not extend the simulation for the dry state at such low frequencies, since the obtained curve shows no trend. This choice is motivated by the computational effort that these simulations require. Simulations are carried out in the NVT ensemble at 300 K. We record stress data every 1 ps. The output data are averaged over the values calculated at every time step. We fit the shear stress ($\tau_{xy}$) with a sinusoidal function:

$$\tau_{xy} = \tau_0 \sin(\omega t + \delta), \tag{5}$$

in which $\delta$ is the delay between the stress and the strain.

According to ref. 83, we define the storage modulus ($G'$) and loss modulus ($G''$) as parameters that give the capability of the material to store (elastic component) and dissipate (viscous component) energy, respectively:

$$G' = \frac{\tau_0}{\gamma_0}\cos\delta \quad G'' = \frac{\tau_0}{\gamma_0}\sin\delta \tag{6}$$

From eqn (3), we estimate tan $\delta$ as:

$$\tan\delta = \frac{G''}{G'}. \tag{7}$$

We organize our outcomes in order to provide a comparison among related systems to highlight how the visco-elastic behavior is affected by: (1) the %HA in the compound; (2) the presence of water; and (3) the amplitude of the angular strain. For each frequency of deformation, we simulate three shear loading cycles and report the average value along with the error bars to highlight the main tendencies.

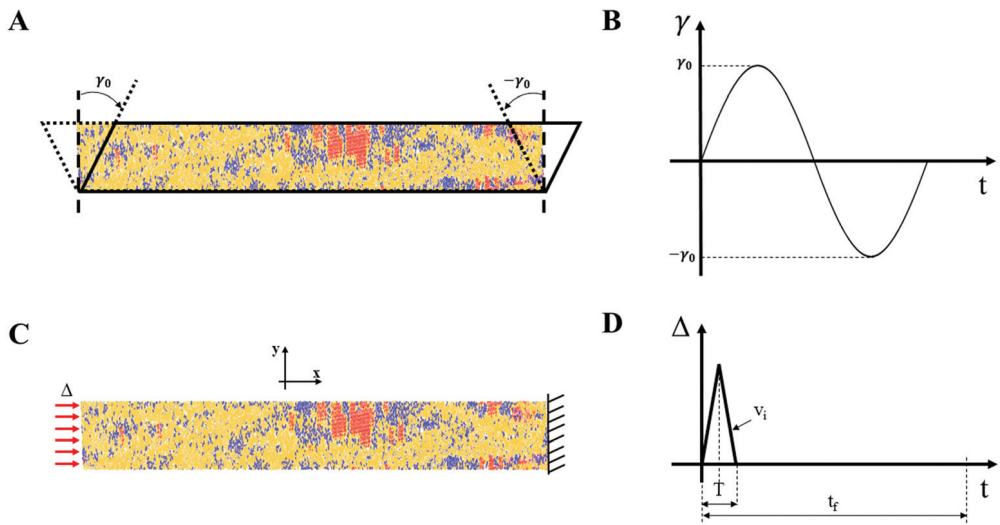

**Fig. 5** Methodologies to study the viscoelastic behavior of the material. (A and B) Shear DMA of the material. We apply a cyclic angular deformation with an amplitude equal to $\gamma_0$ to the full periodic topologies, varying the %HA, amplitude, frequency, and hydration state of the material. (C and D) Wave propagation in the chopped topologies. (C) Topology fixed at one end and loaded by an axial compressive displacement-based load at the opposite end. (D) Impulsive load evolution over time. The load is applied for $T = 2$ ps with an input velocity ($v_i$) equal to 100 m s$^{-1}$, 500 m s$^{-1}$ or 1000 m s$^{-1}$. The observation time $t_f$ is equal to 140 ps.

### Wave propagation analysis (WP)

We fix the atoms belonging to the BD group at one end and use the other to apply a displacement-based load. Based on a previous study that demonstrates the similar outcomes from tensile/compressive loads,[26] we carry out our study using impulsive compressive loads with three reference input velocities ($v_i$): 100 m s$^{-1}$, 500 m s$^{-1}$ and 1000 m s$^{-1}$. The load is delivered in 2 ps, while the total observation time ($t_f$) is 140 ps (Fig. 5C and D).

We post-process the displacements of the $C_\alpha$ atoms along the $x$-axis and estimate the wave speed by taking into account the evolution of the average displacement peak of the traveling wave.

We study the dissipative behavior of the material through the evolution of the square value of the velocity of the wave front ($v_{max}^2$), which resembles the normalized kinetic energy of the system. By assuming the following approximation:

$$v_{max}^2 \sim e^{-\frac{t}{\tau}} \tag{8}$$

we define $\tau$ as the relaxation time of the system, which is the amount of time needed to reduce the amplitude of the function of a factor equal to $e^{-1}$.[26,55,62]

## Author contributions

Conceptualization, M.M., A.D., and M.J.B.; data curation, M.M. and A.D.; formal analysis, A.D.; funding acquisition, M.M. and M.J.B.; investigation, M.M., A.D., S.D., and M.J.B.; methodology, M.M., A.D., G.S.J. and M.J.B.; project administration, M.M. and M.J.B.; resources, M.J.B.; software, M.J.B.; supervision, S.D. and M.J.B.; validation, M.M., A.D., and G.S.J.; visualization, M.M. and A.D.; roles/writing – original draft, M.M.; and writing – review and editing, M.M., A.D., G.S.J., S.D., and M.J.B. All authors intellectually contributed and provided approval for publication.

## Conflicts of interest

All the authors declare no conflict of interest.

## Acknowledgements

This work was supported by the European Union's Horizon 2020 research and innovation program under the Marie Skłodowska-Curie grant agreement COLLHEAR no. 794614. G. S. J. and M. J. B. acknowledge additional support from ONR (N000141612333) and AFOSR (FATE MURI FA9550-15-1-0514), as well as NIH U01HH4977, U01EB014976, and U01EB016422. A. D. was financially supported by the Progetto Roberta Rocca (Doctoral at MIT) Fellowship. G. S. J. acknowledges support from the Laboratory Directed Research and Development (LDRD) Program of Oak